\documentclass[12pt]{iopart}

\def\ApJ{{\it Astrophys. J.}}
\def\MNRAS{{\it Mon. Not. R. Astron. Soc}}
\def\AA{{\it Astron. Astrophys.}}
\def\Nature{{\it Nature}}
\def\GCN{{\it GCN Circ.}}

\usepackage{graphicx}
\usepackage{dcolumn}
\usepackage{bm}
\begin{document}

\title{The Interpretation and Implication of the Afterglow of GRB 060218}
\author{Yi-Zhong Fan$^{1,2,3,4}$, Tsvi Piran$^{1}$, Dong Xu$^{5}$}

\address{1 \ The Racah Inst. of Physics, Hebrew University, Jerusalem 91904, Israel}

\address{2 \ Purple Mountain Observatory, Chinese Academy of
Science, Nanjing 210008, China}

\address{3 \ National Astronomical Observatories, Chinese Academy of
Sciences, Beijing 100012, China}

\address{4 \ Lady Davis Fellow}

\address{5 \ National Dark Cosmology Centre, Niels Bohr Institute, University
of Copenhagen, Juliane Maries Vej 30, 2100 Copenhagen, Denmark}
\ead{yzfan@pmo.ac.cn, tsvi@phys.huji.ac.il, dong@astro.ku.dk}

\begin{abstract}
The nearby GRB 060216/SN 2006aj was an extremely long, weak  and
very soft GRB. It was peculiar in many aspects. We show here that
the X-ray, ultraviolet/optical and radio afterglow of GRB 060218
have to be attributed to different physical processes arising from
different emission regions. From  the several components in this
burst's afterglow only the radio afterglow can be interpreted in
terms of the common external shock model. We infer from the radio
that the blast wave's kinetic energy was $\sim 10^{50}$ erg and
the circumburst matter had a constant rather than a wind profile.
The lack of a ``jet break" up to 22 days implies that the outflow
was wide $\theta_j >1$.  Even though the late X-ray afterglow
decays normally it cannot result from an external shock because of
its very steep spectrum. Furthermore, the implied kinetic energy
would have produced far too much radio. We suggest that this X-ray
afterglow could be attributed to a continued activity of the
central engine that within the collapsar scenario could arise from
fall-back accretion. ``Central engine afterglow" may be common in
under-luminous GRBs where the kinetic energy of the blast wave is
small and the external shock does not dominate over this
component. Such under-luminous GRBs might be very common but they
are rarely recorded because they can be detected only from short
distances.

\end{abstract}



\submitto{Journal of Cosmology and Astroparticle Physics, JCAP}

\maketitle

\section{Introduction}
GRB 060218 \cite{cusu06a} was a  nearby (z=0.033) burst
\cite{mira06, cusu06b} associated with a  bright type Ic
broad-lines SN \cite{modj06, soll06, pian06,mir06,maz06}. It is
distinguished in several aspects from other bursts: (i) It is very
long ($T_{90} \sim 2000$ sec). (ii) The prompt $\gamma-$ray and
X$-$ray luminosity is extremely low $\sim 10^{47}~{\rm
erg~s^{-1}}$ \cite{saka06} and the overall isotropic equivalent
$\gamma$-ray energy, a few $\times 10^{49}$ erg, is small compared
to typical bursts. (iii) The prompt emission is very soft and it
contains a soft thermal component in the X-ray band. The thermal
emission begins at $\sim 152$ sec and continues up to $\sim 0.1$
day. (iv) A second thermal component in the UV/optical band peaks
at $t\sim 1$ day after the GRB trigger \cite{camp06}. (v) For
$t>0.1$ day, the XRT lightcurve is simple and is well described a
single power-law decay $t^{-1.15}$ with no break \cite{camp06}.
(vi) For $t>1.8$ day, the 8.46 GHz radio afterglow lightcurve
decays as $t^{-0.85}$ without break \cite{sod06a}. While the
prompt emission is very different from a typical GRB and the
optical emission is complicated by the appearance of the thermal
bump and the supernova signal this last component, the X-ray and
the radio afterglow, seem to be rather typical.

We focus here on the X-ray and the radio afterglow, and use them
as keys to understand what has happened in this burst. The X-ray
afterglow decays normally and at first\footnote{The first XRT
spectral index $\beta_{_X}$ was quite uncertain. For example, De
Luca \cite{luca06} suggested that $\beta_{_X} \sim -2.3\pm 0.6$,
whereas Cusumano et al. \cite{cusu06b} give later $\beta_{_X} \sim
-2.3\pm 0.2$ and the radio observations \cite{sod06a} were not
available for a while.} one could have interpreted it as arising
from a standard external shock. However, the  X-ray spectrum is
too steep \cite{luca06, cusu06b} to be consistent with this
interpretation. Furthermore, the radio observations \cite{sod06a}
are incompatible with the kinetic energy required to produce the
X-ray emission by an external shock. We suggest that the X-ray
afterglow should be attributed to a ``central engine afterglow"
resulting from a continued activity of the central engine, as
suggested already in 1997 by Katz \& Piran \cite{katz97}. We argue
that such afterglow could be common in the under-luminous nearby
GRBs (see section 2 for details).

The radio afterglow, on the other hand, can be interpreted in
terms of the standard afterglow model. One can infer from it the
kinetic energy, $E_k \sim 10^{50}$ erg, as well as the wide
opening angle, $\theta_j>1$, of the relativistic component of the
ejecta. The association with a type Ic SN suggests that the
progenitor was a WR-star \cite{camp06}. One expects, therefore,
that the central engine is surrounded by a dense stellar wind,
like the one seen in GRB 980425 that was associated with SN 98bw
\cite{li99,waxm04}. However, the density nearest to the progenitor
depends on the mass loss rate during the latest phases of the
WR-star, which is unknown at present \cite{woos03}. With the radio
data, we show that a dense wind profile is not favored (see
section 3 for details).

We examine possible sources for the thermal emission in section 4.
Our conclusions and the implications for the GRB/SN connection are
discussed in section 5.

\section{The long term X-ray emission from the central engine}
The late ($>0.1$ day) X-ray afterglow is similar to the one seen
in typical GRBs in its overall intensity as well as in the almost
standard power law decay index $\alpha_X\sim -1.1$. However, the
time averaged XRT spectral index $\beta_X =-2.2\pm 0.2$ is too
steep to be reproduced in an external shock.   For $\beta_X=-2.2
\pm 0.2$, the power-law distribution index of the shocked
electrons $p=5.4$ or 4.4, depending on the X-ray band being below
or above the cooling frequency $\nu_c$. In the constant density
circumburst medium case, the expected temporal index $\alpha =
(2-3p)/4<-2.8$ for $\nu_X>\nu_c$ otherwise $\alpha =
3(1-p)/4<-2.6$ \cite{sari98, pira99}. In the wind case, the
expected temporal index $\alpha = (2-3p)/4<-2.8$ for $\nu_X>\nu_c$
otherwise $\alpha = (1-3p)/4<-3$ \cite{chev00}. All are far from
consistent with the the observation $\sim -1.1$.

The steep X-ray spectrum enables us to rule out the possibility
that the X-ray emission arises due to inverse Compton. Sari \&
Esin \cite{sari01} have shown that the inverse Compton spectra is
much shallower than $-2.2$ unless it is in the Klein-Nishina
regime. Clearly the observed X-ray cannot be in the Klein-Nishina
regime. Therefore we can rule out the possibility that the X-ray
afterglow arises due to either synchrotron-self inverse Compton
(SSC) or the inverse Compton scattering of the SN optical photons
with the external forward shock electrons.

Even if we ignore the very steep spectrum, the external shock
model is still
 inconsistent because the X-ray emission is strong but the radio emission
 is very weak. Parameters $E_k \sim 10^{51}$erg,
 $\epsilon_e\sim 0.1$, $\epsilon_B \sim 0.01$ and $n\sim 1~{\rm cm^{-3}}$ are
 needed to reproduce the late X-ray emission ($t>0.1$ day). With these parameters
 the resulting radio emission would have been about 1-2 orders brighter than the observation
 \cite{sod06a}.

 An attractive alternative possibility for the production of the X-ray
 afterglow is the a continued activity of the central engine.
 This idea was first proposed by  Katz \& Piran already in 1997 \cite{katz97} and
 had been discussed in the context of GRB 970228  by Katz, Piran \& Sari \cite{katz98}.
 However, the agreement  of the predictions of the external shock afterglow model
 with most subsequent
multi-wavelength afterglows observation and in particular the
smooth light curves
 seen in most afterglow lead to the understanding that afterglows
 are produced by external shocks.
 The energetic soft X-ray flares observe recently in many afterglows of {\it Swift}
 GRBs \cite{brie06} lead  Fan \& Wei \cite{fan05} and  Zhang et al. \cite{zhan06} to
 re-introduce this model and to interpret
 these flares as arising from a continued activity of the central engine.
When proposing the so-called ``late internal shock" model, Fan \&
Wei \cite{fan05} speculated that in some GRBs, the X-ray and
IR/optical afterglow might be attributed to different physical
processes and thus from different regions.  However,  these X-ray
flaring afterglows are quite different from the current
 long term power-law decaying lightcurve. This
 X-ray afterglow of GRB 060218 provides us an indication
 for a power-law decaying afterglow arising from the activity of the
 central engine. Such indications were also seen earlier in some
 pre-Swift GRBs \cite{Bjor02}. If the corresponding outflow is from the central engine
 and there is a significant energy dissipation converting the kinetic energy
 into the X-ray emission, a power-law decaying X-ray central engine afterglow
 should be detected.

 In the following  we call the usual afterglow from
 external shocks a ``fireball afterglow" or
 ``afterglow" and the afterglow attributed to the long lived activity of the
 GRB central engine as a ``central engine afterglow". The
 central engine afterglow, besides those flares detected in {\it
 Swift} GRB X-ray afterglows, are expected
 to be detected in sub-luminous GRBs whose regular afterglows is weak and hence they do
 not over shine this activity. As such sub-luminous GRBs can be
 detected only from relatively short distances we will detect only few such bursts
 even if the total number of such under-luminous bursts is larger than the total
 number of regular GRBs. Alternatively, the ``central engine
 afterglow" component may emerge if (i) The forward shock parameters $\epsilon_e$
 and/or $\epsilon_B$ taken in eq. (\ref{eq:FxAft}) are much smaller than the
 value normalized there; (ii) Some of or all the free
 parameters $\epsilon$, $f_{\rm x}$ and $f_b^{-1}$ taken in eqs.
 (\ref{eq:MacF1}-\ref{eq:FxCen}) have been underestimated
 significantly. In this case, a burst with a ``central engine
afterglow" component may be detectable at high redshift\footnote{A
possible candidate is GRB 060210, a burst at $z\sim 4$
\cite{stan06}. For this burst, the
 R-band flux is just about 10 times that of the
X-ray (at 3.5 keV) and is decaying with time as $t^{-1.3}$ for
$t>500$ s. On the other hand, the XRT spectral index is $-1.17\pm
0.04$ (X. Y. Dai, 2006, private communication). It is thus quite
difficult to interpret these data self-consistently within the
standard external shock model. This inconsistency could be
resolved if the X-ray emission is a ``central engine afterglow"
while the optical emission is the normal external shock
afterglow.}.

To estimate the possible flux from a ``central engine afterglow"
we consider, as an example,  the ``Type II collapsar" model of
MacFadyen et al. \cite{macf01}.
 Clearly if $dM/dt$ that follows
 an under-luminous $\gamma-$ray burst is significantly
 lower than the value taken in Eq.(\ref{eq:MacF1}),
the ``central engine afterglow" emission
 should be dimmer or even undetectable (unless other free
 parameters $\epsilon$, $f_{_X}$, and/or $f_b^{-1}$ taken below are much larger).
 As the difference between the progenitors
 of bright and dim bursts is not clear we assume that
 this accretion rate, which was originally
 suggested for bright bursts, is applicable also for
 sub-luminous ones. Here we take the lowest accretion rate $dM/dt$ presented
in the Fig. 5 of \cite{macf01}:
 \begin{equation}
dM/dt \sim 10^{-6} t_{d,-1}^{-5/3} M_\odot~ {\rm s^{-1}},
\label{eq:MacF1}
 \end{equation}
 where $t_d$ is the observer's time measured in days.
  Here and throughout this text, the convention $Q_x=Q/10^x$ has
  been adopted in cgs units.
 Following MacFadyen et al.  \cite{macf01}, we take an energy conversion
 coefficient $\epsilon \sim 0.001-0.01$ and the beam correction factor
 $f_b \sim 0.01-1$, (note that for this particular burst $f_b \sim 1$)
 thus the outflow
 luminosity can be estimated by
 \begin{equation}
 L\sim \epsilon (dM/dt) c^2/f_b \sim 2\times 10^{46}~{\rm erg~s^{-1}}
 \epsilon_{-3} f_{b,-1}^{-1}t_{d,-1}^{-5/3}.
 \label{eq:MacF2}
 \end{equation}
Assuming that the fraction of the outflow converted into soft
X-ray emission is $f_{\rm x}\sim 0.01-0.1$  and for a luminosity
distance $D_L\sim 10^{27}$ cm, the XRT flux
\begin{eqnarray}
F_{_{\rm X}} &\sim & f_{\rm x} L/(4\pi D_L^2) \nonumber\\
&\sim & 2\times 10^{-10} \epsilon_{-3}f_{\rm x,-1}
f_{b,-1}^{-1}t_{d,-1}^{-5/3}\nonumber\\
&& D_{L,27}^{-2}~~ {\rm erg~s^{-1}~{\rm cm^{-2}}}.
\label{eq:FxCen}
\end{eqnarray}
On the other hand, the forward shock X-ray emission is expected to
be  \cite{fan06a}:
\begin{eqnarray}
F_{_{\rm X}} &\sim&  3\times 10^{-12}~{\rm ergs
~s^{-1}~cm^{-2}}~({1+z})^{\rm (p+2)/4} D_{L,27}^{-2}
\nonumber\\
&& \epsilon_{B,-2}^{\rm (p-2)/4} \epsilon_{e,-1}^{\rm
p-1}\bar{E}_{k,50}^{\rm (p+2)/4}(1+Y)^{-1}t_{d,-1}^{\rm (2-3p)/4},
\label{eq:FxAft}
\end{eqnarray}
where $\bar{E}_k$ is the total isotropic equivalent energy of the
outflow, $\epsilon_e$ and $\epsilon_B$ are the fraction of shock
energy given to the electrons as well as magnetic filed. The
energy $\bar{E}_k$ must include both the energy of the initial GRB
outflow and that of the ``central engine afterglow" outflow $=\int
(1-f_{\rm x}) L dt$ that is added later. As we show later, for GRB
060218 this additional energy is a small fraction (less than 0.1)
of the total energy and its inclusion is insignificant. To get
this and the following numerical coefficients, we take $p=2.3$.
$Y=(-1+\sqrt{1+4\eta \eta_{_{KN}} \epsilon_e/\epsilon_B})/2$ is
the Compoton parameter, where
$\eta=\min\{1,(\nu_m/\nu_c)^{(p-2)/2}\}$ \cite{sari96,wei98},
$0\leq \eta_{_{KN}}\leq 1$ is a coefficient accounting for the
Klein-Nishina effect \cite{fan06a}.

Comparing Eq.(\ref{eq:FxCen}) and (\ref{eq:FxAft}) we see, as one
could expect,  that if the  GRB outflow is significantly less
energetic ($\bar{E}_k \sim E_k \sim 10^{50}$ erg) than typical GRB
($E_k \sim 10^{53}$ erg), the ``central engine afterglow"
component dominates. So the central engine afterglow may be common
for the sub-luminous GRBs. This prediction could be tested in the
coming months or years.

Note that in this particular model the temporal decay ($-5/3$) is
too steep as compared with the observations of $-1.1$. However
this temporal decay (Eq.(\ref{eq:FxCen})) is dictated simply by
the accretion rate used in Eq.(\ref{eq:MacF1}) and surely, there
is enough freedom to allow for  a different slope there.

\section{The late radio afterglow: constraint on the density profile of
the medium}  Multi-wavelength radio data have been presented in
Soderberg et al. \cite{sod06a}. There are 11 detections. 8 of
which are at 8.46 GHz, ranging from $1.8-22$ days. The good
quality 8.46 GHz lightcurve can be well fitted by a single power
law $t^{-0.85}$. This decline slope ($-0.85$) is significantly
different from that of GRB 980425 ($\sim -1.5$). The radio
emission is very weak and has already been discussed when ruling
out the external shock origin for the X-ray afterglow. We examine
it now within the contexts of a constant density and a wind
circumburst medium.

\subsection{A constant density medium }\label{sec:ISM}
For a  constant density medium and for $\nu_m<\nu_a<\nu_{\rm
obs}<\nu_c$, $F_{\nu_{\rm obs}} \propto t^{\rm 3(1-p)/4}$
\cite{sari98, pira99}. To reproduce the current 8.46 GHz
lightcurve $t^{-0.85}$, we need $p\sim 2.1$. The bulk Lorentz
factor of the outflow can be estimated by:
\begin{equation}
\Gamma \approx 3.4E_{k,51}^{1/8}n_0^{-1/8}t_d^{-3/8}(1+z)^{3/8}.
\end{equation}
The lack of a jet break in the radio afterglow up to 22 days after
the burst suggests a very wide jet opening angle $\theta_j>1$
\cite{rhoa99, sari99, halp99}.

In the radio band, the synchrotron self-absorption effect should
be taken into account. Through the standard treatment
\cite{rybi79}, for $\nu_a<\nu_m<\nu_c$, we have
\begin{equation}
\nu_a \approx 1.3\times 10^{10}~{\rm Hz}~
\epsilon_{B,-2}^{1/5}E_{k,51}^{2/5}\epsilon_{e,-1}^{-1}C_p^{-1}(1+z)^{-1}n_0^{3/5}.
\end{equation}
For $\nu_m<\nu_a<\nu_c$, we have
\begin{eqnarray}
\nu_a &\approx & 2.6\times 10^{10}~{\rm Hz}~
\epsilon_{B,-2}^{(p+2)/[2(p+4)]}E_{k,51}^{(p+2)/[2(p+4)]}n_0^{2/(p+4)}\nonumber\\
&&\epsilon_{e,-1}^{2(p-1)/(p+4)}C_p^{2(p-1)/(p+4)}
(1+z)^{(p-6)/[2(p+4)]}\nonumber\\
&&t_d^{-(2+3p)/[2(p+4)]}.
\end{eqnarray}

With the available radio data, we have three constraints on the
physical parameters of the afterglows. One is the self-absorption
frequency $\nu_a \sim 4\times 10^9~{\rm Hz}~>\nu_m$ at $t_d \sim
5$. The other is the 22.5 GHz flux $F(22.5{\rm GHz})\sim 0.25$ mJy
at $t_d\sim 3$. Another is the cooling frequency $\nu_c \leq
5\times 10^{15}$ Hz, which has not been presented in Soderberg et
al. \cite{sod06a} but one can deduce this from their Fig. 2,
provided that the synchrotron spectrum of the external shock
electrons is not dominated in the XRT band. We thus have the
following relations:
\begin{eqnarray}
\epsilon_{B,-2}^{(p+2)/[2(p+4)]}E_{k,51}^{(p+2)/[2(p+4)]}n_0^{2/(p+4)}\nonumber\\
\epsilon_{e,-1}^{2(p-1)/(p+4)} C_p^{2(p-1)/(p+4)}
\sim 0.44,\\
\epsilon_{e,-1}^{\rm p-1}\epsilon_{B,-2}^{\rm (p+1)/4}C_p^{\rm
p-1}
E_{k,51}^{\rm (p+3)/4}n_0^{1/2}\sim 3.2\times 10^{-3},\\
E_{k, 51}^{-1/2}\epsilon_{B,-2}^{-3/2}n_0^{-1}
 \leq 0.43.
\end{eqnarray}
These relations are satisfied with
$(E_k,~\epsilon_e,~\epsilon_B,~n)\sim (10^{50}~{\rm
erg},~0.01,~0.001,~100~{\rm cm^{-3}})$. A similar estimate of
$\epsilon_e$ has also been suggested by Dai, Zhang \& Liang
\cite{dai06} and is it within the range seen in detailed afterglow
modelling of other bursts \cite{pana05}.

\begin{figure}[h!]
\begin{center}
\includegraphics[width=0.58\textwidth,clip=true]{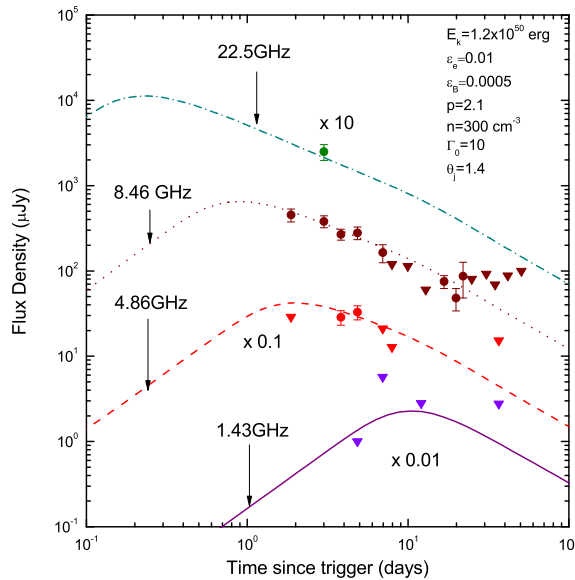}
\caption[...]{A fit to the  observed radio afterglow lightcurves
of GRB 060218 \cite{sod06a} for a constant density circumburst
medium.  The inverted triangles are upper limits ($3\sigma$).
Different colors are for different bands.} \label{fig:ISMFit}
\end{center}
\end{figure}

Fig. \ref{fig:ISMFit} depicts our numerical fit to the radio data.
The code is the same as that used in Fan \& Piran \cite{fan06a}.
One novel effect taken into account is the synchrotron
self-absorption, following the standard treatment \cite{rybi79}.
The external inverse Compton cooling, caused by the long term
X-ray emission from the central engine, has been calculated.
However, this cooling effect changes the current radio emission
only slightly because the inverse Compton parameter $Y_{_{\rm
EIC}}\sim 2L_{\rm
ph,41.5}\epsilon_{B,-3}^{-1}E_{k,50}^{-1}t_{d,1}$ \cite{fan06b}
 is too small to change the distribution of
the shocked electrons significantly (The cooling Lorentz factor is
$\sim 10^6$ and the random Lorentz factor electrons accounting for
the radio afterglow emission is just $\sim$ a few hundreds).
$L_{ph}$ is the luminosity of the X-ray photons from the central
source.

Since we are discussing a relatively late radio afterglow at
$t_d>1.8$, and since the kinetic energy is small the blast wave is
in the sub-relativistic phase at this stage. Therefore the results
do not depend on the initial shape of the ejecta and in particular
the results are insensitive to whether it is thin or thick
\cite{sari95, lazz06}.

So far we have ignored the effects of the injection of additional
energy into the initial GRB blast wave. This happens when the
``central engine afterglow" outflow catches up with the
decelerating initial GRB outflow. The energy injection rate is
$dE_{inj}/dt_d=L(t_d)\sim (1-f_{\rm x})L_X(t_d)/f_{\rm x}$ for
$t>t_o$, where $L_X(t_d)\sim 6.9\times 10^{43}~{\rm
erg~s^{-1}~cm^{-2}}~t_{d,-1}^{-1.15}$ is the observed X-ray
luminosity and $f_{\rm x}$ is conversion efficiency, $t_o$ is the
time at which the ``central engine afterglow" begins. Here we take
$t_o \sim 0.035$ day, i.e, slightly larger than $T_{90}$. For the
prompt emission the conversion efficiency is $ E_\gamma/(E_\gamma
+ E_k)\sim 0.3$, where $E_\gamma \sim 6\times 10^{49}$ erg is the
prompt $\gamma-$ray energy of GRB 060218. With this efficiency the
total energy injected into the initial GRB ejecta is $E_{inj}\sim
\int_{t_o}^{12}L(t_d)dt_d =6.3\times 10^{48}{\rm erg}$. It is much
smaller than the initial kinetic energy $E_k\sim 10^{50}~{\rm
erg}$. Such a weak energy injection cannot influence significantly
the dynamics and emission of the forward shock emission.  Indeed,
detailed numerical calculation taking into account the energy
injection show that the forward shock emission is nearly
unchanged.

A more serious effect is the reverse shock emission that will
arise from the interaction of the ejecta of the late continues
emission with the blast wave formed from the initial outflow. As
an example we consider the reverse shock emission from a baryonic
outflow with a bulk Lorentz factor $\sim \Gamma_o=10$. At $t>0.1$
day, the initial GRB ejecta has been decelerated so the reverse
shock is relativistic.  Since the weak energy injection does not
accelerate the initial GRB ejecta significantly, we can
approximate the Lorentz factor of the reverse shock as
$\gamma_{\rm rs}\approx (1-\beta_o \beta)\Gamma_o\Gamma$, where
$\beta_o=\sqrt{1-1/\Gamma_o^2},~\beta=\sqrt{1-1/\Gamma^2}$. The
magnetic field generated in the reverse shock can be estimated by
$B_{\rm rs}\sim [8\epsilon_B\gamma_{\rm rs}(\gamma_{\rm
rs}-1)L(t)/(\Gamma_0^2 R^2 c)]^{1/2}\approx 0.1~{\rm
Gauss}~\epsilon_{B,-3.3}^{1/2}[\gamma_{\rm rs}(\gamma_{\rm
rs}-1)/12]^{1/2}t_{d,-1}^{-0.58}R_{16.3}^{-1}\Gamma_{o,1}^{-1}$.
The number of electrons in the reverse shock region is $N_{\rm
e,tot}(t)=E_{inj}(t)/(\Gamma_o m_p c^2)\approx 6.1\times
10^{50}\Gamma_{o,1}^{-1}(1.17 - t_{d,-1}^{-0.15})$. The maximum
spectral flux is thus $F_{\rm max,rs}\approx 1.1~{\rm
Jy}~(1+z)N_{\rm e,tot}(t) \Gamma B_{rs}/D_L^2=46~{\rm \mu
Jy}~\epsilon_{B,-3.3}^{1/2}[\gamma_{\rm rs}(\gamma_{\rm
rs}-1)/12]^{1/2} R_{16.3}^{-1}\Gamma_{o,1}^{-2}\Gamma
t_{d,-1}^{-0.58}(1.17 - t_{d,-1}^{-0.15})$ \cite{sari98, fan05}.
For these parameters  $F_{\rm max,rs}<100~{\rm \mu Jy}$ at $t \sim
0.1$ day and decreases with time continually. This flux is much
lower than the optical and radio observations (Note that the
reverse shock emission flux in any bands is $\leq F_{\rm
max,rs}$). The reverse shock X-ray flux is also far below the
observation. We thus conclude that the radio afterglow is
dominated by the forward shock and the model is self-consistent.

\subsection{A circumburst  wind }
For a stellar wind \cite{dai98, mesz98}, $n=3\times 10^{35}A_*
R^{-2}~{\rm cm^{-3}}$, where $A_*=[\dot{M}/10^{-5}M_\odot~{\rm
yr^{-1}}][v_w/(10^8{\rm cm}~{\rm s^{-1}})]$ \cite{chev00},
$\dot{M}$ is the mass loss rate of the progenitor, $v_w$ is the
velocity of the wind.

In the relativistic regime the bulk Lorentz factor of the ejecta
$\Gamma^{\rm w}\propto t^{-1/4}$, where the superscript ``${\rm
w}$" represents the wind model. Following Chevalier \& Li
\cite{chev00}, we have the maximal spectrum flux $F_{\nu,{\rm
max}}^{\rm w} \propto t^{-1/2}$, $\nu_m^{\rm w}\approx 1.3\times
10^{10}~{\rm Hz}~\epsilon_{e,-1}^2 C_p^2 E_{k,50}^{1\over
2}\epsilon_{B,-2}^{1\over 2}(1+z)^{1\over 2}t_d^{-3/2}$ and
$\nu_c^{\rm w} \approx 3.2\times 10^{13}~{\rm
Hz}~\epsilon_{B,-2}^{-3/2}E_{k,50}^{1\over
2}A_*^{-2}(1+z)^{-3/2}t_d^{1\over 2}$, where $C_p \equiv
13(p-2)/[3(p-1)]$. For $\nu_m^{\rm w}<\nu_a^{\rm w}<\nu_{\rm
obs}<\nu_c^{\rm w}$ (where $\nu_{\rm obs}=8.46$ GHz is the
observer frequency), as suggested by the observation at $t_d>1.8$,
the observed lightcurve is
\begin{equation}
F_{\nu_{\rm obs}}= F_{\nu,{\rm max}}^{\rm w} (\nu_{\rm
obs}/\nu_m^{\rm w})^{(p-1)/2} \propto t^{\rm (1-3p)/4}.
\label{eq:Fr1}
\end{equation}

In the Newtonian regime the velocity of the ejecta satisfies
$\beta \propto t^{-1/3}$, the radius of the shock front $R\propto
t^{2/3}$, the magnetic field strength $B \propto \beta R^{-1}
\propto t^{-1}$. Furthermore, we have $F_{\nu,{\rm max}}^{\rm
w}\propto R B \propto t^{-1/3}$, $\nu_m^{\rm w}\propto
\beta^4B\propto t^{-7/3}$, $\nu_c^{\rm w}\propto t$. Therefore,
for $\nu_m^{\rm w}<\nu_a^{\rm w}<\nu_{\rm obs}<\nu_c^{\rm w}$, we
have
\begin{equation}
F_{\nu_{\rm obs}} \propto t^{\rm (5-7p)/6}. \label{eq:Fr2}
\end{equation}
For $p\geq 2$, the resulting temporal indexes are $\leq -1.25$ and
$\leq -1.5$, (for Eqs. (\ref{eq:Fr1}) and \ref{eq:Fr2}
respectively) are  much steeper than the observed slope of
$-0.85$.

While the $p<2$ possibility cannot be ruled out but it is less
likely as particles accelerated at relativistic shocks usually
have a power law distribution index $p\geq 2$ \cite{gall02}. On
the other hand, as shown earlier, in a constant density medium a
$p\sim 2.1$ can reproduce the data quite well. So we conclude that
a dense wind model is less likely.

An important question is whether  the energy injection caused by
the late activity of the central engine can  flatten the
lightcurve significantly and thus render the wind profile likely?
The answer is negative. Consider an energy injection with the form
$dE_{inj}/dt \propto  (t/t_o)^{-q}$,  where $q$ is a constant. In
this case $q\sim 0.4$ is needed to flatten the afterglow
lightcurve at $t\gg t_o$ significantly  (see Table 2 of
\cite{zhan06} and the references therein and the detailed
numerical calculation of \cite{fan06a}). However, the X-ray light
curve requires   $q \sim 1.15$. For such a large $q$ and as the
total energy of this outflow is small compared to the initial
energy, the energy injection cannot modify the temporal behavior
of the forward shock emission in such an energy injection.
Similar to the constant density medium case, it is straightforward
to show that the corresponding reverse shock emission in radio
band is unable to account for the data.  We  thus are left with
the conclusion that the wind profile is unlikely.

\section{The thermal emission}
A soft thermal component is seen \cite{camp06, lian06} in the
X-ray spectrum comprising $\sim 20\%$ of the 0.3-10 keV flux. It
begins at $\sim 152$ sec and lasts up to $\sim 10^4$ sec. The
fitted black body temperature shows a marginal decrease ($kT\simeq
0.16-0.17$ keV, with $k$ the Boltzmann constant) and a clear
increase in luminosity, by a factor of 4 in the time range
300s-2600s, corresponding to an increase in the apparent emission
radius from $R_{_{\rm BB,XRT}}=(5.2\pm 0.5)\times 10^{11}$ cm to
$R_{_{\rm BB,XRT}}=(1.2\pm0.1)\times 10^{12}$ cm \cite{camp06}. In
the sharp decline phase, the XRT emission is dominated by a
thermal component ($kT=0.10\pm 0.05$ keV, the corresponding
apparent emission radius is $R_{_{\rm
BB,XRT}}=6.6^{+14}_{-4.4}\times 10^{11}$ cm). This thermal
component is undetectable in later XRT observation.

A second  thermal component is detected by the UVOT. At 1.4 days
$\sim 120$ ksec the black body peak is centered within the UVOT
passband. The fitted values are $kT=3.7^{+1.9}_{-0.9}$ eV and
$R_{_{\rm BB,UVOT}}=3.29^{+0.94}_{-0.93}\times 10^{14}$ cm,
implying an expansion speed of $(2.7\pm 0.8)\times 10^9~{\rm
cm~s^{-1}}$. This speed is typical for a supernova and it is also
comparable with the line broadening observed in the optical
spectra \cite{pian06}. The UVOT thermal component is therefore
very likely dominated by the expanded hot stellar envelope (see
also Campana et al. \cite{camp06}).

The nature of the X-ray thermal emission is less clear.  Campana
et al. \cite{camp06} suggest that it  arises from a shock break
out from a very dense wind ($A_*>30$) surrounding the SN
progenitor. As we have shown earlier the medium surrounding the
progenitor is not likely to be a dense wind, as  required in this
model \cite{camp06}. We suggest, therefore, that the XRT thermal
component arises from a shock heated stellar matter. As the size
of the emitting black body region ($6 \times 10^{11} - 10^{12}$
cm) is larger than the size of a typical WR- star ($10^{11}$ cm)
there are two possibilities: The emission could be from a
 a hot cocoon surrounding the GRB
ejecta \cite{rami02, zhan04} and expanding initially with $v \sim
0.1 c$. An alternative possibility is that the X-ray thermal
emission arises from the shock break out from the stellar
envelope. This would require, however, a progenitor's size of
$\sim 10^{12}$ cm (see also \cite{li06}). This is much larger than
$\sim 10^{11}$ cm or less, that is expected from a star stripped
from its H, He and probably O, as inferred from the spectroscopic
analysis of the SNe \cite{pian06}. It is not clear if stellar
evolution model can accommodate such a progenitor, but surprises
of this nature have happened in the past. A relativistic
radiation-hydrodynamics calculations are needed to test the
viability of these two possibilities. This is beyond the scope of
this work.

Here we just show that a hot and optical thick outflow could
account for the temporal behavior of the XRT and UVOT thermal
emission. After the central engines turns off (i.e., there is no
fresh hot material injected), the hot outflow expands and cools
adiabatically as $T\propto n_p^{1/3}\propto R^{-\alpha/3}$, where
$\alpha=3$ if the hot outflow is spreading and  $\alpha=2$
otherwise, $n_p$ is the number density of the particle. Once the
hot region cools adiabatically so that $kT \ll 0.2 $ the thermal
emission recorded by XRT in the range 0.2 to 10 keV  decrease
quickly with time as
\begin{equation}
L_{\rm th,XRT} \propto R^2 e^{{\rm -0.2keV}/kT}\propto R^2
e^{-\alpha R/3R_0},
\end{equation}
where $R_0$ is the radius of the outflow at the turning off time
of the central engine. The V-band flux is $L_{\rm th,V}\approx
4\pi \sigma R^2 T^4 {y^3 \Delta y \over e^y-1}$, where $y=2.3{\rm
eV}/kT$, $\Delta y\approx 0.13y$, accounting for the FWHM width of
V-band. For $y\ll 1$,
\begin{equation}
L_{\rm th,V} \propto TR^2 \propto R^{2-\alpha/3}\propto
t^{2-\alpha/3},
\end{equation}
increases with time until $y \sim 1$ and then it decreases
rapidly. As noted by Campana et al. \cite{camp06}, such a behavior
is in agreement with {\it Swift}'s observations.

\section{Discussion and Summary}

The recent nearby burst GRB 060218 had many peculiar features.
There are several  components of the observed afterglow, X-ray
optical and radio and there is no simple afterglow model that can
fit two out of the three. From these components only the observed
radio afterglow at $t \sim 10^5$ sec is rather usual and its
lightcurve is compatible with a weak burst with a low kinetic
energy. We summarize the situation below:

\begin{itemize}

\item The temporal decay of the bright (non-thermal) X-ray
afterglow ($t>10^4$ s) looks like typical. However, the very steep
spectrum and the very weak radio afterglow rule out an external
shock origin (both form  synchrotron radiation and from inverse
Compton emission). We suggest following the earlier suggestion of
Katz \& Piran \cite{katz97} and Katz, Piran \& Sari \cite{katz98}
that this emission arises due to continued activity of the central
engine. This is the first time that a power-law decaying X-ray
afterglow is attributed to the activity of the central engine,
though it has been suggested by Fan \& Wei \cite{fan05} and Zhang
et al. \cite{zhan06} that the flare-rich X-ray afterglow that have
been detected in a good fraction of {\it Swift} GRBs \cite{brie06}
also trace the long term activity of the central engine.

\item The radio afterglow can be understood within the standard
blast wave  model, provided that the medium is ISM-like, the
overall kinetic energy is $10^{50}$erg, and the fraction of shock
energy given to the electrons is $\sim 0.01$. A wind profile is
disfavored as it requires $p\sim 1.5$ and even then it is not
clear if one can reproduce the observation.

\item The lack of a ``jet break" of 8.46 GHz afterglow lightcurve
up to 22 day indicates that the outflow is very wide ($\theta_j
>1$). This is somewhat at odds with the standard Collapsar model
that involves a narrow jet.

\item The X-ray and optical/UV thermal emission cannot arise from
the relativistic ejecta. A shock heated envelope of the progenitor
is the most natural source. The question whether the envelope has
expanded rapidly or was it initially large is open.

\end{itemize}

There are several implications to these conclusions. First we note
that in the current event, the long term power-law decaying X-ray
afterglow, the ultraviolet/optical afterglow and the radio
afterglow cannot be attributed to the same physical process and
they arise from different regions. While this is sort of expected
for the thermal optical and X-ray components it is somewhat
puzzling and alarming that the two nonthermal X-ray and radio
components do not seem to arise from the same source. While it may
indicate a serious problem in the overall fireball model we
suggest that this is a manifestation of the fact that GRBs are
much more complex than was anticipated earlier and that indeed
different components such as external shocks as well as continued
activity of the central engine might take place simultaneously. Of
course GRB 060218 is not unique in this case and such a complexity
might have been seen in other peculiar multi-wavelength afterglows
which have been poorly interpreted within the standard external
shock model \cite{fan06a}.

We have argued that  there is  a clear indication that  a
power-law decaying, non-thermal X-ray afterglow that cannot be
produced by an  external shock  and we have shown in \S{2} that
fall-back accretion within a collapsar might be energetic enough
to power a detectable ``central engine afterglow". This component
is in particular important for the sub-luminous GRBs, for which
the ejecta is significantly less energetic than that of the
typical GRBs and the late activity of the central engine is not
hidden by the external shock afterglow. Such sub-luminous GRBs can
be detected only from small distances. As it is possible and even
likely (Nakar 2006, private communication) that the real rate of
such GRBs is significantly higher than the rate of regular GRBs
they should dominate the nearby bursts population. We thus predict
that such central engine afterglow would be detected in a good
fraction of nearby GRBs. In fact with a slight change of the
paraments it is possible that the central engine afterglow might
dominate over the external shock afterglow even for brighter GRBs.
This, for example, could arise in GRB 060210.

The power-law decaying ``central engine afterglow" of GRB 060218
was identified by its very steep X-ray spectrum. However, the
inconsistency of the strong X-ray afterglow with the weak radio
signal was essential to verify this idea. In general the X-ray
spectrum of the central engine afterglow is not necessarily so
steep. For example, within the well detected X-ray flares, that
have been attributed to continues activity of the central engine,
just a few have a very steep spectrum (see Table 8 of
\cite{bult06}). Therefore, a multi-wavelength afterglow analysis
is essential to identify the ``central engine afterglow"
component.

Finally we mention two observed features that seem to be
inconsistent with the canonical Collapsar model. The first among
the two is the lack of a clear wind profile. The afterglows arises
at a distance  of $R\sim 10^{17}$ cm from the central engine. It
could be that the observed profile arose from the interaction of
the wind with the surrounding matter or it may reflect a low
mass-loss rate of the progenitor star during the post-helium
burning phases. A similar feature was seen also in many GRBs but
here we have information on regions that are nearer to the central
engine. The wide angle of the relativistic ejecta is also
incompatible with the usual Collapsar model, in which a narrow jet
punches a narrow hole in the envelope of a WR star \cite{zhan04}.
This may indicate that GRB 060218 was an almost ``failed GRB". Due
to some unique feature of the progenitor (a larger than usual
size? or a smaller than typical mass?) the relativistic ejecta
almost did not make it across the envelope. This has lead to a
wide relativistic outflow with an unusually low initial Lorentz
factor. This, in turn, lead to the softer spectrum (possibly due
to internal shocks taking place in a region with optical depth of
order unity). A significant fraction of the energy was given to a
hot cocoon and was reprocessed as a thermal emission - seen both
in X-ray and later in the UV/optical. One can speculate that in
many other cases the relativistic ejecta would have stopped
completely and we would have a ``failed GRB". It is possible that
this is the reason why GRBs are not seen in most type Ib, c SNe
\cite{berg03, sode06}.

\ack We thank E. Waxman for helpful discussions and an anonymous
referee for constructive comments. This work is supported by
US-Israel BSF and by the ISF via the Israel center for High Energy
Astrophysics. TP acknowledges the support of Schwartzmann
University Chair. YZF is also supported by the National Natural
Science Foundation (grants 10225314 and 10233010) of China, and
the National 973 Project on Fundamental Researches of China
(NKBRSF G19990754). DX is at the Dark Cosmology Center funded by
The Danish National Research Foundation.

\end{document}